\documentclass[aip,jcp,reprint]{revtex4-1}
\usepackage{graphicx}
\usepackage{amsmath}
\usepackage{amssymb}
\usepackage{mhchem}

\begin{document}
\title{Ions at Hydrophobic Interfaces}

\author{Yan Levin}
\email{levin@if.ufrgs.br}
\affiliation{Instituto de F\'isica, Universidade Federal do Rio Grande do Sul, Caixa Postal 15051, CEP 91501-970, Porto Alegre, RS, Brazil}

\author{Alexandre P. dos Santos}
\email{alexandre.pereira@ufrgs.br}
\affiliation{Departamento de F\'isica, Universidade Federal de Santa Catarina, 88040-900, Florian\'opolis, Santa Catarina, Brazil}

\begin{abstract}

We review the present understanding of the behavior of ions at the 
air-water and oil-water interfaces.  We  argue that while the alkali metal cations remain strongly hydrated 
and are repelled from the hydrophobic surfaces, 
the anions must be classified into kosmotropes and chaotropes.  The 
kosmotropes remain strongly hydrated in the vicinity of a hydrophobic surface, while the 
chaotropes loose their hydration shell and can become adsorbed to the interface.  The mechanism
of adsorption is still a subject of debate.  Here, we argue that there are two driving forces for 
anionic adsorption:  
the hydrophobic cavitational energy and the interfacial electrostatic surface potential of water. While the
cavitational contribution to ionic adsorption is now well accepted, the role of the electrostatic 
surface potential is much less clear.  The difficulty is that even the sign of this potential is a subject
of debate, with the \textit{ab initio} and the classical force fields simulations predicting electrostatic
surface potentials of
opposite sign.  In this paper, we will argue that the strong anionic adsorption found in the polarizable force
field simulations is the result of the artificial electrostatic surface potential present in the classical
water models. We will show that if the adsorption of anions would be as large as predicted by 
the polarizable force field simulations,
the excess surface tension of \ce{NaI} solution would be strongly negative, 
contrary to the experimental measurements. While the large polarizability of heavy halides is a fundamental
property and must be included in realistic modeling of the electrolytes solutions, 
we argue that the point charge water models, studied 
so far, are incompatible with the polarizable ionic force fields
when the translational symmetry is broken. The goal for the future should be the development of water
models with very low electrostatic surface potential.  We believe that such water 
models will be compatible with the polarizable force fields and can then be used to study the interaction
of ions with hydrophobic surfaces and proteins.

\end{abstract}

\maketitle

\section{Introduction}

Availability of highly reactive halogen ions at the surface of aerosols has tremendous implications for the atmospheric chemistry~\cite{HuSh95,KnLa00,ClDo07}. Yet, neither simulations, experiments, nor existing theories are able to provide a fully consistent description of the electrolyte-air interface. The state of the art simulations can be divided into two categories - classical polarizable force fields molecular dynamics (PFFMD)~\cite{PeBe91,DaSm93,StBe99,JuTo01,JuTo02,JuTo06,BrDa08,HoHe09} and \textit{ab initio} quantum density functional theory (DFT)~\cite{ToJu01,KaKu08,KaKu09,Le10} simulations. Both of these simulation methods find that large strongly polarizable halide anions, such as Bromide and Iodide, can become adsorbed at the air-water interface. In fact, classical force fields predict a very strong adsorption of \ce{I-} at the interface. If the degree of adsorption would be as large as predicted by the classical simulations, addition of \ce{NaI} to water would lead to lowering of the surface 
tension of the air-water interface, similar to what happens in solutions containing surfactants. This, however, is not what is found 
experimentally. 
Addition of 
salt to water increases the surface tension of the air-water interface~\cite{WePu96,MaTs01}. Presently, available computational resources do not allow for large scale \textit{ab initio} simulations. Nevertheless, it is now possible to simulate a water slab containing about 200 water molecules and one Iodide ion. The \textit{ab initio} simulations can be used to calculate the potential of mean force (PMF) for the interaction of \ce{I-} with the interface. This potential is significantly less atractive than the one obtained in the classical simulations~\cite{ToSt13}. The PMF calculated using quantum DFT~\cite{BaMu11} was found to be in almost quantitative agreement with the recently introduced polarizable anion dielectric continuum theory~(PA-DCT)~\cite{Le09,LeDo09}. In this review, we will explore the mechanisms which drive the adsorption of highly polarizable ions to the air-water interface and to other hydrophobic surfaces. We will use the PA-DCT to show that the strong adsorption found using PFFMD simulations 
arises from the artificial electrostatic surface potential present in the classical water 
models.
The geometry of these models leads to a surface dipole layer  --- the interfacial water molecules are oriented so that one of the partially charged hydrogens sticks out into air.  This  results in an electrostatic potential drop of approximately $600$~mV across the air-water interface, with air being more electropositive than water~\cite{SoTi97,ZaBr97}. On the other hand, the \textit{ab initio} simulations find a potential drop of the opposite sign - quantum mechanics predicts that the electronic cloulds of water molecules spill out into air, making air more electronegative than bulk water. We will argue that the artificial surface potential produced by the classical water models is the driving force behind the excessive anionic adsorption found using PFFMD simulations. Therefore, unless these models are modified to remove the artificial surface potential, they can not be used to study ions at the air-water 
interface or any other hydrophobic surface.

\section{Hydrophobic Interfaces}

The study of electrolytes at aqueous interfaces is a classical problem of  physical chemistry, going back over a century ago to the pioneering works of Gibbs, Langmuir~\cite{La17}, Wagner~\cite{Wa24}, and Onsager and Samaras~\cite{OnSa34}. In spite of this long and venerable history, the interaction of ions with hydrophobic  interfaces is still poorly understood and remains a  subject of a great  debate~\cite{VeAl13,NeWe13,MoNe13,DoLe13c,Ge13,ToSt13}. The behavior of ions at interfaces is of great practical importance in such diverse fields as atmospheric chemistry, electrochemistry, colloidal science, biophysics, physical chemistry, etc.  In the case of atmospheric chemistry, it is important to know how ions are distributed inside sea-salt aerosols, since the presence of highly reactive halogens at the surface of microscopic  water droplets can lead to production of acid rain and the destruction of tropospheric ozone~\cite{HuSh95,KnLa00,ClDo07}. Adsorption of ions to hydrophobic residues can lead to denaturation 
of proteins
and can affect colloidal stability. Over a hundred years ago, Hofmeister observed that there is a significant degree of specificity in the interaction of ions with proteins  --- while addition of some salts can lead to precipitation of protein solutions, other ions can make solutions more stable.  The effect of salt on proteins is much more sensitive to anion than to cation.  The Hofmeister (lyotropic) series has now been observed in many different systems and  has been found to  affect  micellar formation~\cite{JiLi04,Ro07,LuMa11,MuMo12,MuDe13,MaKa13}, bacterial growth~\cite{LoNi05}, ionic liquids~\cite{AoKi13,WaSu13}, liquid crystals~\cite{DaLa09,CaMa12}, microemulsions~\cite{MuMo04}, critical coagulation concentrations of  colloidal suspensions, ~\cite{LoJo03,LoSa08,PeOr10,DoLe11,CaFa11,ScNe12}, etc.   
Over a century ago~\cite{He10}, the lyotropic series was also observed in the surface tension measurements of electrolyte solutions~\cite{AvSa76,WePu96,MaTs01,LiDe13}.   An explanation for why salts increase the surface tension of the air-water interface was provided by Wagner~\cite{Wa24} and Onsager and Samaras~\cite{OnSa34} (WOS), who argued that as an ion approaches a dielectric interface, it induces a surface charge which repels  it from the interface.  On the basis of the Gibbs adsorption isotherm equation WOS then argued that ionic depletion from the interfacial region will result in increased surface tension.   Contrary to this explanation, however, recent photoelectron-spectroscopy measurements~\cite{MaGi91,MaPo94,Ga04,Gh05} have shown that some anions can be present at the air-water interface. Subsequently,
atomistic molecular dynamics simulations~\cite{PeBe91,DaSm93,StBe99,JuTo01,JuTo02,JuTo06,BrDa08,HoHe09} and quantum \textit{ab initio} simulations~\cite{ToJu01,KaKu08,KaKu09,Le10} have confirmed that some polarizable anions might be adsorbed to the air-water interface. The ionic propensity for hydrophobic surfaces was, once again, found to follow the Hofmeister series~\cite{Ho88,MeBa12,RePa12}, showing that WOS theory is incomplete. Dispersion 
interactions, neglected in the WOS approach, were suggested to be responsible for the ionic specificity~\cite{BoWi01}.  It was soon realized, however, that although the dispersion forces are  important for the interaction of ions with  the
hydrophobic surfaces,
they can not explain the propensity of large halogen anions for the  air-water interface  ---  
the dispersion interactions favor the
adsorption of small weakly polarizable cations and not of strongly polarizable anions~\cite{MaRu03}.

Recently, a new theory was developed which allows us to calculate the surface tensions for different electrolytes at various hydrophobic interfaces~\cite{Le09,LeDo09,DoDi10,DoLe10,DoLe12}.  The results of the theory are in excellent agreement with the experiments. 
The theory shows that while the alkali metal cations remain strongly hydrated 
and are repelled from the hydrophobic surfaces, 
the anions belong to two categories: kosmotropes and  chaotropes.  The structure making
kosmotropes remain strongly hydrated in the vicinity of a hydrophobic surface, while the structure breaking
chaotropes loose their hydration shell and can become adsorbed to the interface.  It is important to stress that, in this theory, the notion of kosmotropes and chaotropes has nothing to do with the long-range influence of ions on the hydrogen bond network of water, instead, it only refers to the local ionic hydration.  
The theory allows us to 
explore in great detail the various driving forces responsible for the ionic adsorption to hydrophobic surfaces.  
The interaction potentials predicted by the theory can be
compared with the PMF obtained
using the explicit-water molecular dynamics simulations, 
and the effect of the ion-interface interaction on the thermodynamics properties of electrolyte solution --- such as its
surface tension  --- can be easily calculated.  

At the moment, there is an intense debate on the role
of the interfacial electrostatic surface potential of water on ionic
adsorption~\cite{ArBe12,BaSt12}. The classical point charge water models predict a surface potential of a neat air-water interface to be approximately $-600$~mV~\cite{SoTi97,ZaBr97}, while the potential obtained using \textit{ab initio} quantum DFT simulations is of the opposite sign and is significantly larger, $+3000$~mV~\cite{KaKu08,KaKu09}.  Nevertheless, when 
properly coarse grained,
the surface potential of  \textit{ab initio} simulations vanishes,  while the classical surface  potential  persists~\cite{KaKu11}. In this
paper, we will show that the electrostatic surface potential of SPC/E water is partially responsible for the excessive
adsorption predicted by the polarizable force field simulations.  Furthermore, we will show that if the adsorption of anions
is as strong as predicted by these simulations, the excess surface tension of \ce{NaI} would be strongly negative, instead of positive, as  measured experimentally. Below we will review the PA-DCT and explore the contributions of cavitation, polarizability, dispersion and the electrostatic
surface potential on ion-interface interaction.

\section{Theory: Water-Air Interface}

We, first, briefly review the interaction potential between an ion and the air-water interface~\cite{Le09,LeDo09,DoDi10}. The potential is constructed by taking into account the polarization, hydration, cavitation and image charges. This potential will then be used in a modified  Poisson-Boltzmann equation  to calculate the ionic density distribution near an interface. 

The standard model of  electrolyte solutions threats ions as hard spheres with a point charge located at the center. This is the basis of the celebrated Debye-H\"uckel~(DH)~\cite{DeHu23} theory.  This theory, and its subsequent extensions, such as the Mean Spherical Approximation~(MSA) and the Hypernetted Chain Equation~(HNC), have been found to be very accurate for describing bulk properties of electrolyte solutions~\cite{HaMc86}. On the other hand, rigid charge distribution of DH and Onsager~and~Samaras~\cite{OnSa34} theories does not permit ionic presence at the air-water interface.  The reason for this is that the electrostatic 
self-energy penalty for exposing the rigid ionic charge to the low dielectric environment overwhelms any other entropic or enthalpic gain in free energy, arising from the surface solvation~\cite{TaCo10}.  
Ionic polarizability appears to be the key ingredient necessary to understand ionic adsorption at the hydrophobic interfaces~\cite{Le09}.  Polarizable ions can shift their electronic charge density so that it remains mostly hydrated by the water molecules of the topmost interfacial layer.  Large polarizability decreases dramatically the self-energy penalty of surface solvation.

A polarizable ion can be modeled as an imperfect spherical conductor of relative polarizability $\alpha=\gamma_i/a^3$, where $\gamma_i$ is the ionic polarizability. Note that for a perfect conductor $\alpha=1$. The electrostatic self energy of an ion at distance $z$ from the interface can be written as~\cite{Le09}
\begin{eqnarray}\label{Upol}
\beta U_{p}(z;x)=\left\{
\begin{array}{l}
\frac{\lambda_B}{2 a} \text{ for } z\ge a \ , \\
\frac{\lambda_B}{2 a}\left[\frac{\pi x^2}{\theta(z)}+\frac{\pi
(1-x)^2 \epsilon_w}{[\pi-\theta(z)]\epsilon_o}\right] + \\
g \left[x-\frac{1-cos[\theta(z)]}{2} \right]^2 \text{ for } -a<z<a  \ ,
\end{array}
\right.
\end{eqnarray}
where $\lambda_B=\beta q^2/\epsilon_w$ is the Bjerrum length, $7.2$~\AA\ for water at room temperature, $g=(1-\alpha)/\alpha$, $\theta(z)=\arccos(-z/a)$ and $x$ is the fraction of the ionic charge that remains hydrated. Minimizing Eq.~\eqref{Upol}, we obtain the expression for $x(z)$,
\begin{equation}
x(z)=\dfrac{\dfrac{\lambda_B \pi \epsilon_w}{a \epsilon_o
\left[\pi-\theta(z)\right]}+g [1-cos[\theta(z)]]}{\dfrac{\lambda_B
\pi}{a \theta(z)} + \dfrac{\lambda_B \pi \epsilon_w}{a \epsilon_o [\pi-\theta(z)]} +2 g} \ .
\end{equation}
Inserting this expression into Eq.~\eqref{Upol}, yields the electrostatic self energy of an ion at distance $z$ from the interface, $U_p(z)$.  We find that the self energy of a polarizable ion located at the interface is an order of magnitude smaller than the energy of a hard non-polarizable ion at the same position~\cite{TaCo10}. From the electrostatic perspective, therefore, polarizable ions will still prefer bulk solvation.

What then drives polarizable ions towards the interface?
To solvate an ion, requires creation of a cavity into which the ion is inserted. It is clear that any perturbation to the
hydrogen bond network of water costs energy which, for small cavities, is predominantly entropic and scales with the volume of the cavity.
Clearly, if ion is expelled from water it will results in a cavitational free energy gain.  We will, therefore, suppose that the cavitational free energy is proportional to the ionic volume exposed to the aqueous medium~\cite{RaTr05,Le09}. The cavitation potential can then be written as
\begin{eqnarray}\label{cavpot}
\beta U_c(z)=\left\{
\begin{array}{l}
 \nu a^3 \text{ for } z \ge  a  \ , \\
 \frac{1}{4} \nu a^3  \left(\frac{z}{a}+1\right)^2 \left(2-\frac{z}{a}\right)
\text{ for } -a<z<a \ ,
\end{array}
\right.
\end{eqnarray}
where $\nu=0.3/$\AA$^3$ is obtained using SPC/E water simulations~\cite{RaTr05}.

The broken translational symmetry, imposed by the air-water interface, leads to two additional contribution to the ion-interface 
interaction free energy.  As the ion approaches the interface, the spherically symmetric screening of its electric field is perturbed, resulting in higher electrostatic energy.  The dielectric discontinuity also results in the  build up of the surface charge, both of these effects lead to the repulsion of  the ion from the interface.  The energy cost to bring an ion from bulk water  to a distance $z$ from the air-water interface was calculated by Levin and Flores-Mena~\cite{LeFl01}. The Levin-Flores-Mena potential can be well approximated by~\cite{DoDi10, DoLe13b}
\begin{eqnarray}\label{Uim}
\beta U_{i}(z)=\left\{
\begin{array}{l}
\beta Wa\dfrac{e^{-2 \kappa (z-a)}}{z} \ \text{ for } z \ge  a  \ , \\
\beta W \dfrac{z}{a} \text{ for } 0 \le z< a \ , \\
0 \text{ for } -a \le z <  0 \ ,
\end{array}
\right.
\end{eqnarray}
where $\kappa=\sqrt{8 \pi \lambda_B c_s}$ is the inverse Debye length. The potential at contact, $W$, is given by~\cite{DoLe13b},
\begin{equation}\label{image_potential_i}
\beta W=\frac{\lambda_B}{2} \int_0^\infty dk \ \frac{k\ f_1(k)}{p\ f_2(k)}  \ ,
\end{equation}
where 
\begin{eqnarray*}
f_1(k)=p \cosh{(k a)}-k \sinh{(k a)}+\nonumber \\
\frac{\epsilon_o}{\epsilon_w}p\sinh{(k a)}-\frac{\epsilon_o}{\epsilon_w}k \cosh{(k a)} \ ,
\end{eqnarray*}
\begin{eqnarray*}
f_2(k)=p \cosh{(k a)}+k \sinh{(k a)} +\nonumber \\
\frac{\epsilon_o}{\epsilon_w}p\sinh{(k a)}+\frac{\epsilon_o}{\epsilon_w}k \cosh{(k a)} \ ,
\end{eqnarray*}
and $p=\sqrt{k^2+\kappa^2}$.

A somewhat different approach for calculating the ion-interface surface potential has been recently proposed by Wang and Wang~\cite{WaWa13}.  For the air-water or oil-water interfaces, the two models lead to very similar interaction potentials.

\section{Surface Tension: The Drop Model}

\begin{figure}[h]
\vspace{0.2cm}
\includegraphics[width=8.cm]{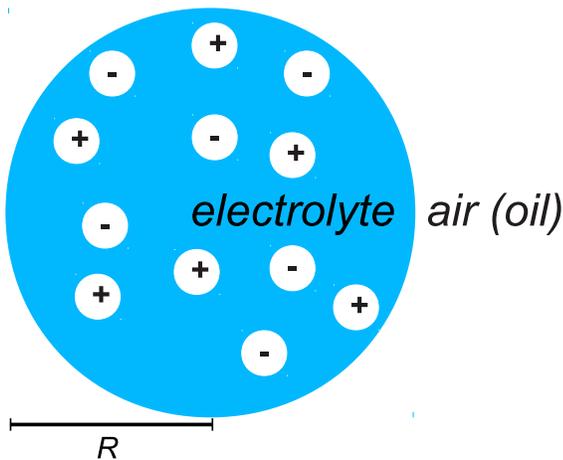}\vspace{0.2cm}
\caption{Illustration of the spherical drop model.}
\label{fig_drop}
\end{figure}

The ion-interface interaction potentials derived in the previous section can be used to calculate the excess surface tension of an electrolyte-air interface. Consider cations and anions of radii $a$ inside a water drop of radius $R$, see Fig~\ref{fig_drop}. For simplicity of notation, we use the same letter $a$ to denote the radii of all ions, note, however, that the numerical value of $a$ will be different
for each ion. Furthermore, while for kosmotropes $a$ is the hydrated radius~\cite{Ni59}, for chaoptropes it is the crystallographic (Latimer) radius~\cite{LaPi39}. The chaotropic ions can cross the interface up to the maximum distance $r_m=R+a$ from the center of the drop. We do not need to consider larger distances, since the growing self energy makes it very improbable for an ion to move farther than this into the low dielectric phase.
The Kosmotropic ions remain strongly hydrated and reach, at most, the distance $r_m=R-a$ from the center of the drop.  The drop is taken to be sufficiently large, $R=300~$\AA, so that the curvature effects can be neglected.
The interfacial tension is calculated using the Gibbs adsorption~(GA) isotherm equation,
\begin{equation}\label{gae}
{\rm d} \gamma=-\Gamma_+ {\rm d} \mu_+ - \Gamma_- {\rm d} \mu_- \ ,
\end{equation}
where $\Gamma_{\pm}=\left[N - V \rho_{\pm}(0) \right]/S$ are the ionic excess, $\mu_\pm$ are the chemical potentials, $N$ is the number of cations or anions, $\rho_{\pm}(0)$ are the bulk concentrations, and $S$ and $V$ are the surface and the volume of the drop, respectively.  The bulk concentrations are obtained from the numerical solution, in spherical coordinates, of the modified Poisson-Boltzmann~(PB) equation:
\begin{eqnarray}\label{pb}
\nonumber
\nabla^2 \phi(r)&=&-\frac{4\pi q }{\epsilon_w} \left[\rho_+(r)-\rho_-(r)\right]  \ ,  \\
\rho_{\pm}(r)&=& A_{\pm} e^{\mp \beta q \phi(r)-\beta U_{\pm}(r)} \ ,  \\
\nonumber
A_{\pm}&=&N\left[4\pi \int_0^{r_
m} dr\ r^2 e^{\mp \beta q \phi(r)-\beta U_{\pm}(r)}\right]^{-1} \ ,
\end{eqnarray}
where $r$ is the distance from the center of the drop, $q$ is the proton charge, $\phi(r)$ is the electrostatic potential with $\phi(0)=0$ and $\phi'(0)=0$, and $\rho_\pm(r)$ are the ionic density profiles. At the level of PB theory, correlations between the ions are ignored, which has been found to be a very reasonable approximation for 1:1 aqueous electrolytes  with concentrations up to 1M, considered in this paper~\cite{Le02,DiDo12}. The ionic chemical potentials, $\beta \mu_\pm= \log{[\Lambda_\pm^3 \rho_{\pm}(0)]}$, are uniform throughout the drop, where $\Lambda_\pm$ are the thermal de Broglie wavelengths. The Gibbs dividing surface~(GDS) is defined at $r=R$. The GDS separates the aqueous and the vapor mediums, modeled as continuum uniform dielectrics with constants $\epsilon_w$ and $\epsilon_o$, respectively. The heterogeneity of the dielectric constants is taken into account by the ion-interface interaction potentials, $U_\pm(r)$,  discussed in the previous section and defined later in the text.

With the ionic potentials in hand, we can solve the PB equation, Eq.~\eqref{pb}, iteratively and calculate the excess surface tensions of various electrolyte solutions, Eq.~\eqref{gae}. For halides, the separation of ions into kosmotropes and chaotropes has been found to be very strongly correlated with their crystallographic radii -- small halogen ions, such as \ce{F-} and \ce{Cl-} produce very intense electric fields which interact strongly with the surrounding water molecules leading to formation of ion-water complexes.  On the other hand,  in the vicinity of a fluctuating interface, a relatively weak electric field of large halides is not sufficient to keep water molecules bound to them, so that chaotropes can become ``dehydrated''. Unfortunately, for polyatomic anions, there is no
simple correlation between either ionic size or ionic polarizability and the ionic hydration.   
However, we find that there is an excellent correlation~\cite{DoDi10} between the Jones-Dole (JD) viscosity $B$ coefficient~\cite{JeMa95,Ma09} and ionic hydration near a hydrophobic interface. 
The JD $B$ coefficient is obtained from a phenomenological fit of the excess viscosity produced by the addition of salt to water,
\begin{equation}\label{jde}
\eta_r=1+A \sqrt{c}+B c \ ,
\end{equation}
where $\eta_r$ is the relative viscosity, $c$ is the concentration of electrolyte, and $A$ and $B$ are the fitting parameters, obtained experimentally. The coefficient $A$ is due to the relaxation of ionic atmosphere perturbed by the shear flow and, for small concentrations, can be calculated using the Debye-H\"uckel-Onsager-Falkenhagen theory~\cite{HaOw58}. On the other hand, the $B$ coefficient depends on the microscopic ion-solvent interaction. For structure making ions (kosmotropes), the $B$ coefficient is positive, while for  structure breaking ions (chaotropes) it is negative.  It is curious that a dynamical property, such as viscosity,
is found to be so strongly correlated with a static property, such as the interfacial tension of the interface.  
The reason for this correlation might be that the strong fluctuations of the instantaneous 
interface can strip the weakly bound water molecules from a chaotropic ion, similar to what happens in a shear flow.
Unfortunately, at the moment, there is no quantitative theory of ionic hydration, for which quantum effects seem to be important.  To understand the difficulties involved,  
consider the iodate ion, \ce{IO3-}.   This ion is very large and strongly polarizable 
(polarizability of $8~$\AA$^3$).  On the other hand, its positive viscosity $B$ coefficient, see Table \ref{tab1},  classifies it as a kosmotrope.  In spite of its large size and polarizability it should, therefore, remain strongly hydrated near the air-water interface.  This should be contrasted with the smaller and less strongly polarizable iodide, \ce{I-}, which is a chaotrope and must loose its hydration sheath near
a hydrophobic surface.  It is impossible to understand this dichotomy simply on the basis of ionic size and polarizability, both of which suggest that \ce{IO3-} should be an excellent chaotrope. The recent \textit{ab initio} simulations~\cite{BaPh11} show a very curious electronic structure of \ce{IO3-}, which might explain the peculiar hydration properties of iodate. In the absence of a theory of ionic hydration we will, therefore, adopt the viscosity $B$ coefficient as an indicator of kosmotropic/chaotropic ionic classification.  
Since kosmotropes remain hydrated, for these ions we will impose a hard-core repulsion
at one hydrated radius from the interface. On the other hand, the chaotropes loose their hydration sheath
so that for these ions, we will need their bare radii as an input for our theory.  The hydrated radii are taken from  Nightingale~\cite{Ni59} and the bare radii from Latimer~{\it et al.}~\cite{LaPi39}. In Table~\ref{tab1}, we summarize the ionic classification  and the parameters used in the potentials.
\begin{table}[b]
\small
\centering
\caption{Ion classification into chaotropes (c) and  kosmotropes (k). Effective radii (hydrated or partially hydrated) for kosmotropes and (bare) for chaotropes.  For chaotropes  we have also included the ionic polarizabilities, which are irrelevant for kaotropes. The polarizabilities are taken from the Ref.~\cite{PyPi92}, the bare radii from Ref.~\cite{LaPi39} and the hydrated radii from Ref.~\cite{Ni59}.}
\label{tab1}
\begin{tabular}{c|c|c|c} 
      \hline
      \hline      
      Ions        &    chao/kosmo     & radius (\AA)   &  polarizability (\AA$^3$) \\
      \hline
      \hline      
      \ce{F-}     &     k             &        3.52    &         --         \\
      \hline
      \ce{Cl-}    &     k             &        2       &         --         \\
      \hline      
      \ce{Br-}    &     c             &        2.05    &        5.07        \\
      \hline      
      \ce{I-}     &     c             &        2.26    &        7.4         \\
      \hline      
      \ce{IO3-}   &     k             &        3.74    &         --         \\
      \hline       
      \ce{BrO3-}  &     k             &        2.41    &         --         \\
      \hline      
      \ce{NO3-}   &     c             &        1.98    &        4.09        \\
      \hline      
      \ce{ClO3-}  &     c             &        2.16    &        5.3         \\
      \hline
      \ce{ClO4-}  &     c             &        2.83    &        5.45        \\
      \hline
      \ce{CO3 -2}  &     k             &        3.94    &         --         \\
      \hline
      \ce{SO4 -2}  &     k             &        3.79    &         --         \\
   \end{tabular}
\end{table}

\begin{figure}[b]
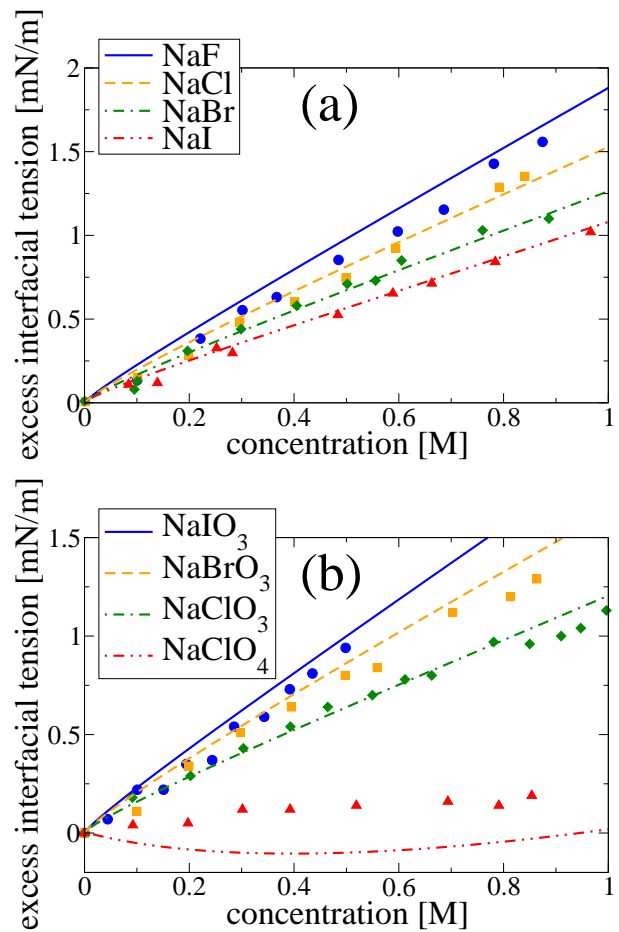

\vspace{0.2cm}
\includegraphics[width=8.cm]{fig_1_ten_a.eps}\vspace{0.2cm}
\includegraphics[width=8.cm]{fig_1_ten_b.eps}\vspace{0.2cm}
\caption{Excess interfacial tensions as a function of salt concentration. The theory is represented by the lines, while the symbols are the experimental data~\cite{MaTs01,MaMa99,Ma}. In panel~(a) the circles, squares, diamonds, and triangles represent the experimental data for salts \ce{NaF}, \ce{NaCl}, \ce{NaBr} and \ce{NaI}, respectively. In panel~(b) the circles, squares, diamonds, and triangles represent the experimental data for salts \ce{NaIO3}, \ce{NaBrO3}, \ce{NaClO3} and \ce{NaClO4}, respectively.}
\label{fig_salt}
\end{figure}

The anion-interface interaction potentials are $U_-(z)=U_i(z)$ for kosmotropes, and $U_-(z)=U_{p}(z)+U_c(z)+U_i(z)$ for chaotropes. 
The potential for cation \ce{Na+}, which is also a kosmotrope,  is $U_+(z)=U_i(z)$. 
Note that $z$ is the distance from the interface, while $r$, in the PB equation, is the distance from the center of the drop.  
Starting with 
\ce{NaI} solution, we adjust the ionic hydration radius of  \ce{Na+} to obtain the best fit of the experimental  surface tension data, see Fig.~\ref{fig_salt}. We find that $a=2.5$~\AA $\,$ results in an excellent agreement with experiment. 
The same ionic radius of \ce{Na+} is then used to calculate the surface tensions of all other sodium salts.  
We see that the theory agrees very well with all the experimental data, except for \ce{NaClO4}, see Fig.~\ref{fig_salt}. 
The deviation from the experimental data, in this case, might be due to an overestimate of the effective radius of \ce{ClO4-}.
Since the cavitational energy scales with the cube of the ionic radius, a small error can result in a significant
overestimate of ionic adsorption. 
An excellent agreement between the theory and experiment for \ce{NaIO3} shows that, in spite of its huge size and large polarizability, iodate remains strongly hydrated and is repelled from the air-water interface.

The theory can also be used to calculate the excess electrostatic potential difference across the air-water interface, resulting from the preferential
anion adsorption, $\Delta\phi=\phi(\infty)-\phi(0)$. Frumkin~\cite{Fr24} was the first to measure a negative value of $\Delta\phi$, showing that there is some partitioning of ions across the interface. In Table~\ref{tab2}, we present the theoretical results for the excess electrostatic potentials of various $1$~M electrolyte solutions and compare them with the experimental measurements of Frumkin~\cite{Fr24} and Jarvis and Scheiman~\cite{JaSc68}.
In spite of a large scatter in the experimental data, there is a reasonable qualitative agreement between the theory and experiments.
\begin{table}[t]
\small
\centering
\caption{Experimental and calculated electrostatic surface potential differences for $1$~M electrolytes.}
\label{tab2}
\begin{tabular}{c|c|c|c} 
      \hline
      \hline      
       Salts       &    Calculated~(mV)   &   Ref.~\cite{Fr24,Ra63}~(mV) & Ref.~\cite{JaSc68}~(mV)  \\
      \hline
      \hline      
      \ce{NaF}     &      4.7             &        --       &         --         \\
      \hline
      \ce{NaCl}    &     -2.1             &        -1       &     $\approx$ -1   \\
      \hline      
      \ce{NaBr}    &     -9.4             &        --       &     $\approx$ -5   \\
      \hline      
      \ce{NaI}     &     -14.3            &       -39       &     $\approx$ -21  \\
      \hline      
      \ce{NaIO3}   &      5               &       --        &       --           \\
      \hline      
      \ce{NaBrO3}  &     -0.12            &       --        &       --           \\
      \hline      
      \ce{NaNO3}   &      -8.27           &       -17       &     $\approx$  -8  \\
      \hline      
      \ce{NaClO3}  &     -11.02           &       -41       &       --           \\
      \hline
      \ce{NaClO4}  &     -31.1            &       -57       &       --           \\
      \hline
      \ce{Na2CO3}  &      10.54           &         3       &     $\approx$   6  \\
      \hline
      \ce{Na2SO4}  &      10.17           &         3       &     $\approx$   35 \\
   \end{tabular}
\end{table}

The ion-interface interaction potential for \ce{I-} calculated using the PA-DCT theory, $U_-(z)=U_p(z)+U_c(z)+U_i(z)$, and the  PMFs calculated using the PFFMD and the {\it ab initio} DFT simulations are plotted in the panel~(a) of Fig.~\ref{fig_pmfs}. The agreement between the interaction potential calculated using the PA-DCT theory and the PMF of the {\it ab initio} simulation is evident. On the other hand, the PMF of a classical PFFMD simulation has a potential well of $\approx$ $3~k_BT$, significantly larger than what is seen in the {\it ab initio} simulation.  In the following section we will explore the origin of this discrepancy. 

\section{Surface Potential of Water}

\begin{figure}[h]
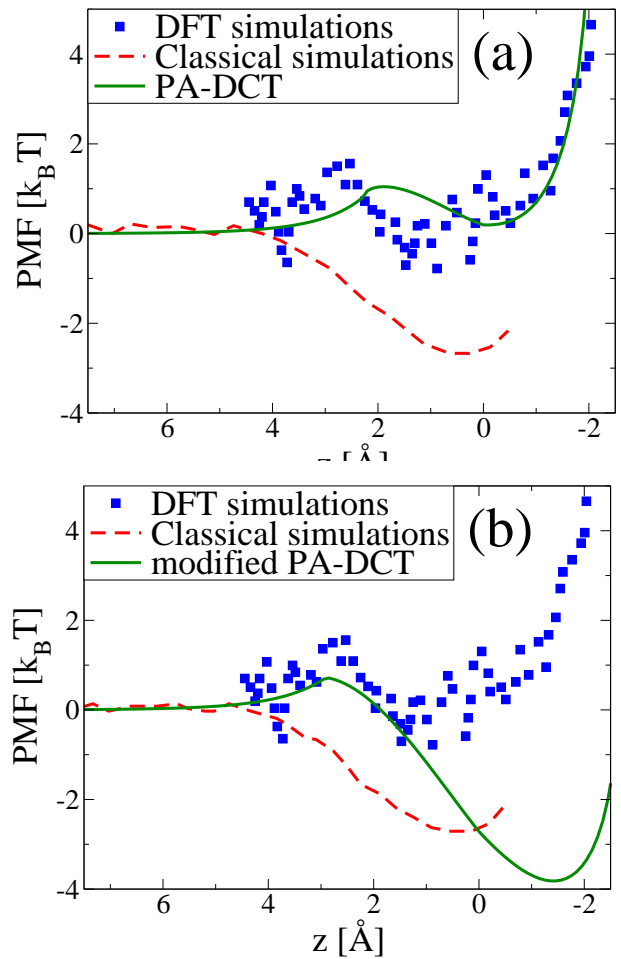

\includegraphics[width=8cm]{fig_pmf_a.eps}\hspace{0.1cm}
\includegraphics[width=8cm]{fig_pmf_b.eps}\hspace{0.1cm}
\caption{Comparison between of the ion-interface interaction potentials  obtained from the pure and dipole-layer modified PA-DCT, and the PMFs calculated using DFT \textit{ab initio} and the classical PFFMD simulations. In panel~(a), the comparison is with the pure PA-DCT theory, in panel~(b), with the dipole-layer modified PA-DCT one. The simulation data is taken from the Ref.~\cite{ToSt13}}.
\label{fig_pmfs}
\end{figure}

The PA-DCT theory shows an excellent agreement with the experimental measurements of surface tensions of  various
electrolyte solutions, with only one adjustable parameter, the radius of \ce{Na+}. This suggests that
the ion-interface interaction potentials predicted by this theory are quite accurate. Yet, when compared with the PMF of \ce{I-} calculated using PFFMD simulations, there is a dramatic difference, see panel~(a) of Fig.~\ref{fig_pmfs}.
While the PA-DCT predicts only a small metastable minimum for \ce{I-}
adsorption at the air-water interface, the PFFMD finds a global minimum of almost $3k_B T$ deep! Clearly, 
existence of such strong attractive interaction between  \ce{I-} and the interface would result in a strong adsorption. Indeed, as we will show later, if the adsorption would be as strong as predicted by the 
PFFMD, the excess surface tension of \ce{NaI} solution would be strongly negative, contrary to the experimental measurements.  Furthermore, in agreement with the  PA-DCT and contrary to PFFMD,
the quantum DFT \textit{ab initio} simulations also find only a small metastable minimum for the PMF of \ce{I-}.
The question of fundamental importance is then: What produces such 
strong attraction between anions and the air-water interface in the PFFMD?
The huge number of parameters in classical water and ion models makes it very difficult to 
untangle the different contributions to free energy and to 
attribute cause and effect~\cite{ArBe12}.  Here, we will argue that the driving
force for the excessive anionic adsorption in these models
comes partially from the artificial electrostatic surface potential. 
The geometry of the classical point charge water models leads to a dipole layer resulting from a broken translational symmetry. To optimize the hydrogen bond network, the surface water molecules become 
oriented so that one of the partially charged hydrogens sticks out into air.  
This  leads to a dipole layer with an excess of positive charge in air and negative charge in water.  
For SPC/E water model, the  dipole layer results in an electrostatic potential drop of approximately $550$~mV
across the air-water interface, with air being more electropositive than water~\cite{SoTi97,ZaBr97}.
To understand the effect of electrostatic surface potential on ionic adsorption of polarizable ions, 
we can include this contribution into our PA-DCT.     
To do this we add to the polarization energy, Eq.~\eqref{Upol} (in the region $-a<z<a$),
a contribution due to 
the interaction of ionic electronic charge with the artificial dipole water layer. 
Recalling that $x$ is the fraction of the ionic charge that is solvated, the gain in electrostatic energy due
to the interaction of an anion with the water dipole-layer is then $q\Delta\chi[1-x(z)]$, where 
$\Delta\chi=-550$~mV, is the electrostatic surface potential of SPC/E water~\cite{SoTi97,ZaBr97}.
The polarization potential, Eq.~\eqref{Upol} (with dipole-layer ion contribution), $U_{p\chi}(z;x)$, must then be minimized to calculate the ionic charge that remains hydrated as the ion moves across the interface, $x(z)$. Substituting $x(z)$ back into  $U_{p\chi}(z;x(z))$, we obtain the interaction energy of an ion with the dielectric dipole-layer interface.  

To make a direct comparison with the PFFMD, we need to make a few additional modifications.  Unlike the real
water with relative dielectric permittivity of $80$, SPC/E water has a dielectric constant of about $69$~\cite{VaVa98,BaSt12}.  Furthermore, unlike the ions of PA-DCT,  ions of PFFMD are not hard spheres, so there is no simple mapping between the radii used in PA-DCT and the Lennard-Jones parameters of the PFFMD.  Our strategy will then be to adjust the effective ``hard-core" radius of iodide ion to get the same adsorption as found in the PFFMD simulations~\cite{JuTo02}.

To compare with PFFMD, we solve the PB equation, Eq.~\eqref{pb}, in the slab geometry, 
with the modified potential, $U_-(z)=U_{p\chi}(z)+U_i(z)+U_c(z)$. The concentration of \ce{NaI} is taken to be $1.2$~M, the same as used in PFFMD~\cite{JuTo02}. The Bjerrum length is modified to $\lambda_B=8.34$~\AA, to account for the reduced dielectric constant of SPC/E water. 

\begin{figure}[t]
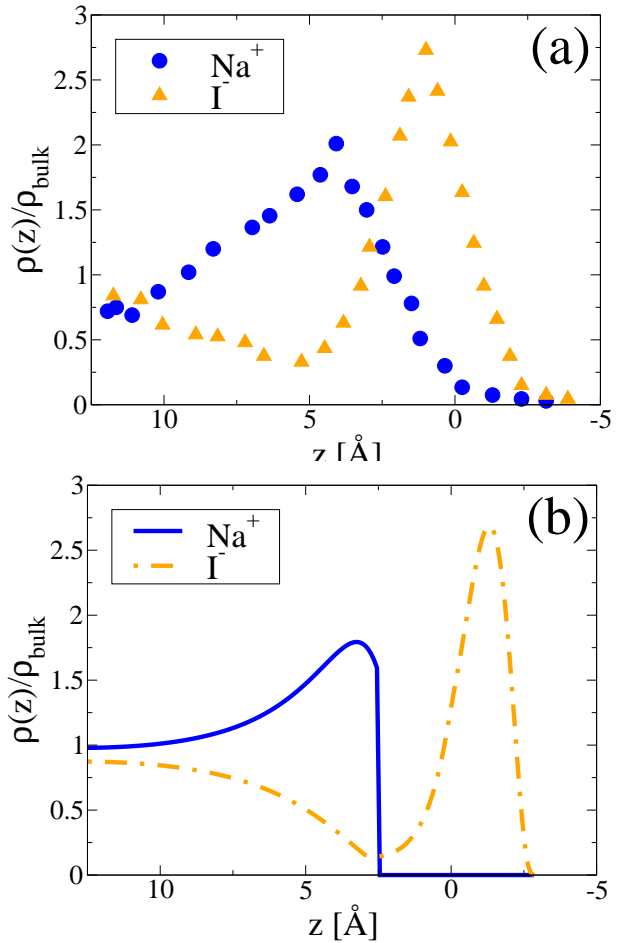

\includegraphics[width=8cm]{fig_NaI_a.eps}\hspace{0.1cm}
\includegraphics[width=8cm]{fig_NaI_b.eps}\hspace{0.1cm}
\caption{Comparison between the density profiles obtained using PFFMD simulations (symbols) and the dipole-layer modified PA-DCT (lines), for the \ce{NaI} salt. The simulation data are obtained from Fig.~$10$ of Ref.~\cite{JuTo02}.}
\label{fig_NaI}
\end{figure}
\begin{figure}[h]
\includegraphics[width=8.cm]{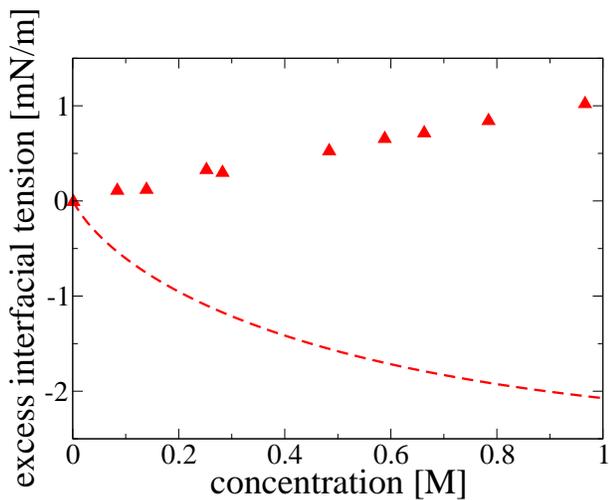}\vspace{0.2cm}
\caption{Excess interfacial tension as a function of salt concentration for \ce{NaI} salt. The dipole-layer modified PA-DCT  is represented by the line, while symbols are the experimental data.}
\label{fig_salt_bad}
\end{figure}

We find that to get the same adsorption of iodide, as seen in PFFMD, using our dipole-layer modified PA-DCT, 
the hard-core radius of \ce{I-} must be changed to $a=2.9$~\AA, instead of the Latimer radius of $a=2.26$~\AA, see Fig~\ref{fig_NaI}.  However, after this modification, the density profiles both of \ce{Na+}
and \ce{I-} become very similar to the ones observed in the PFFMD simulations.  We note, however, that 
due to the hard core repulsion from the interface 
imposed by the PA-DCT on the kosmotropic ions, our model
predicts slightly less adsorption of sodium, than is found in PFFMD simulations.  
Integrating the Gibbs adsorption isotherm equation, we can now calculate the excess surface tension of the \ce{NaI}
solution.  As expected, the excess surface tension of the dipole-layer modified PA-DCT is strongly negative,
contrary to the experimental data, see Fig.~\ref{fig_salt_bad}. It is important to stress that, because of a 
stronger adsorption of \ce{Na+} observed in PFFMD simulations, see Fig~\ref{fig_NaI}, 
the surface tension of \ce{NaI} in classical point charge water models~\cite{DoMu08} will be even lower (more negative) 
than is predicted by the dipole-layer modified PA-DCT.  This is clearly incorrect, showing 
that there is a fundamental incompatibility of polarizable force fields with the currently used  point charge water models.

In panel~(b) of Fig.~\ref{fig_pmfs}, we show the comparison of the ion-interface interaction potential calculated using the dipole-layer modified PA-DCT and the PMF calculated using PFFMD simulations. The two potentials are very similar, although the theoretical result has a slightly larger minimum and is shifted more towards the vapor phase.  The discrepancy  might be due to the fact that the  PMF of the simulations 
is plotted with respect to the GDS and not with respect to the instantaneous interface~\cite{WiCh10}.  
Indeed, when the simulation PMF is replotted
with respect to the instantaneous interface~\cite{StBa13}, the two potentials become very similar.  

The PA-DCT theory has been shown to accurately predict the surface tensions of electrolyte solutions.  It has
also provided us with valuable insight into the origin of the excessive anionic adsorption observed in PFFMD simulations.  In the following sections, we will use the PA-DCT to study the surface tensions of acids~\cite{DoLe10} and the interfacial tensions of electrolyte-oil interfaces~\cite{DoLe12}.

\section{Acid Solutions}

\begin{figure}[h]
\vspace{0.2cm}
\includegraphics[width=8.cm]{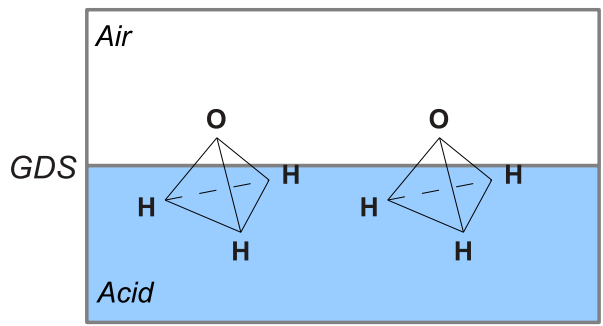}\vspace{0.2cm}
\caption{Illustration of hydronium ions adsorbed at the acid-air interface. Observe the preferential orientation of  hydronium.}
\label{fig_hydronium}
\end{figure}

\begin{figure}[t]
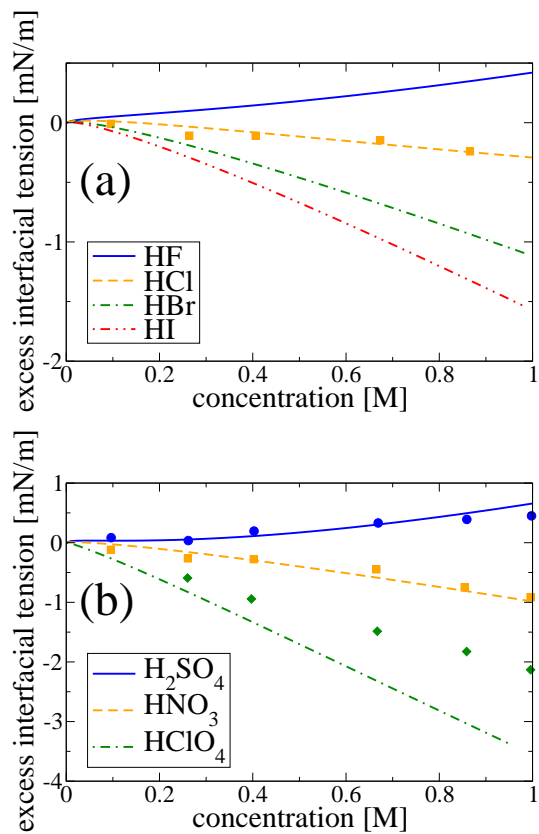

\vspace{0.2cm}
\includegraphics[width=7.cm]{fig_2_ten_a.eps}\vspace{0.2cm}
\includegraphics[width=7.cm]{fig_2_ten_b.eps}\vspace{0.2cm}
\caption{Excess interfacial tensions as a function of salt concentration. Our theory is represented by the lines, while the symbols are the experimental data~\cite{WePu96}. In panel~(a) the circles represent experimental data for acid \ce{HCl}. In panel~(b) the circles, squares and diamonds represent experimental data for acids \ce{H2SO4}, \ce{HNO3} and \ce{HClO4}, respectively.}
\label{fig_acid}
\end{figure}

Unlike salts, addition of most acids to water causes a decrease of the surface tension of solution-air interface~\cite{RaSc66,WePu96}.  It is well known that a proton \ce{H+} interacts strongly with water molecules, resulting in formation of complexes~\cite{Ei64,MaTu99,Zu00} such as \ce{H3O+} and \ce{H2O5+}. In particular, the hydronium ion, \ce{H3O+}, has a piramidal trigonal structure with hydrogens at the base and oxygen at the top~\cite{MuFr05}. In this geometry, oxygen is a bad hydrogen bond receptor, while hydrogens are good hydrogen bonds donors~\cite{PeIy04}, providing an amphiphilic character to \ce{H3O+} behaviour~\cite{Da03,PeSa05}, see Fig.~\ref{fig_hydronium}.
Quantum {\it ab initio} simulations~\cite{PeIy04,TaMa11}, experiments~\cite{PeSa05}, and classical simulations~\cite{MuFr05} all indicate large surface activity of hydronium ion. 
It is straightforward to modify the PA-DCT to explore the thermodynamics of  acid solutions.  
The amphiphilic character of hydronium
results in a strong adsorption of this ion at the air-water interface.  The interaction of 
hydronium with the interface can be modeled by an attractive square well potential.  
The range of the potential  is taken to be one hydrogen bond length, $1.97$~\AA, 
from the interface. The depth of the potential is then adjusted to fit the surface tensions of one of the electrolyte solutions, i.e.  
\ce{HCl}, see Fig.~\ref{fig_acid}.  The same hydronium-interface potential
is then used to calculate the surface tensions of all other acids. 
For ions, the interaction potentials are the same as the ones used in the previous sections of this Review.
The ionic and hydronium density profiles are calculated by solving the modified PB equation, Eq.~\eqref{pb}, with the hydronium potential given by $U_+(z)=U_h(z)+U_i(z)$, where $U_h(z)$ is the square well potential and $U_i(z)$ is the charge-image interaction potential with zero radius, $a=0$. The excess surface tension of \ce{HCl} solution 
is then calculated by integrating the Gibbs adsorption isotherm, Eq.~\eqref{gae}.  We find that if the depth of the square well potential is adjusted to  $-3.05~ k_BT$, we obtain an excellent agreement with the experimental surface tensions of the hydrochloric acid, Fig.~\ref{fig_acid}.  The same hydronium potential,
\begin{eqnarray}\label{uhtotal}
\beta U_h(z)=\left\{
\begin{array}{l}
0 \, \text{ for } z  \ge  1.97\ \text{\AA} \ , \\
-3.05 \, \text{ for } 0 \le z < 1.97\ \text{\AA} \ ,
\end{array}
\right.
\end{eqnarray}
is then used to calculate the surface tension of all other acids, Fig.~\ref{fig_acid}. 
Although for most acids we find a very good agreement with the experimental measurements,  significant deviations are observed for \ce{HClO4}.  This is similar to what has been found for \ce{NaClO4} and is, again, attributed to the overestimate of the effective radius of \ce{ClO4-}, quoted in the literature.

\section{Water-Oil interface}

The hydrophobic surfaces are more complicated than the air-water interface.  Besides the cavitational energy
responsible for the chaotropic  adsorption at the air-water interface, we must also consider possible dispersive
interactions with the low-dielectric hydrophobic medium. At this time there are no {\it ab initio} simulations
indicating the value of the electrostatic potential difference across the neat water-hydrophobe interface.  In the
absence of such measurements, we will suppose that, similar to the air-water interface, the electrostatic surface 
potential is negligible. 

To construct a theory of ionic interaction with a hydrophobic surface  --- which will be modeled as an interface between water and oil --- we will follow the same procedure developed for the air-water interface.  
We will suppose that kosmotropic ions will remain hydrated and will be repelled from the interface, while the chaotropic ions can loose their hydration sheath and become adsorbed at the interface. Since the dielectric constant of oil is very low, we can use the same
polarization and charge-image potentials, Eq.~\eqref{Upol} and \eqref{Uim}, respectively, developed for the air-water interface. 
The cavitational energy gain, however, is different for a fluid-fluid interface than for the air-water interface.  As the ion moves from water to oil, it decreases the perturbation to aqueous environment, while increasing perturbation to oil. For small cavities, this energy is mostly entropic. 
The molecular weight of dodecane, used in experiments, is 10 times larger than that of water, while its mass density is very similar to water, so that the number of oil molecules excluded from a cavity of radius $a$ is, on average, 
an order of magnitude smaller than of water molecules for a cavity of the same radius. This means that 
the cavitational penalty of creating a hole in oil should be an order of magnitude lower than for 
creating a cavity of the same radius in water. Therefore, for high molecular weight oils, the cavitational
potential, Eq.~\eqref{cavpot}, will be the same as for the air-water interface~\cite{DoLe12}.

\begin{figure}[t]
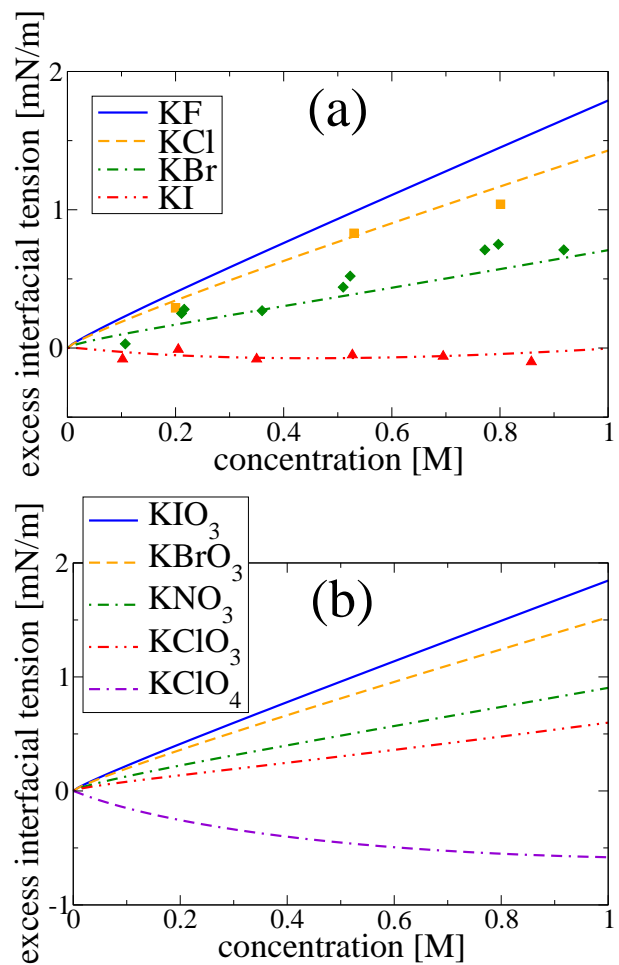

\vspace{0.2cm}
\includegraphics[width=8.cm]{fig_3_ten_a.eps}\vspace{0.2cm}
\includegraphics[width=8.cm]{fig_3_ten_b.eps}\vspace{0.2cm}
\caption{Excess interfacial tensions as a function of salt concentration. The theory is represented by the lines, while the symbols are the experimental data~\cite{AvSa76}. In panel~(a) the squares, diamonds and triangles represent experimental data for salts \ce{KCl}, \ce{KBr}, and \ce{KI}, respectively. There is no experimental data for salts represented in the panel~(b).}
\label{fig_salt_oil}
\end{figure}
The dispersion interaction between ions and oil arises from the quantum fluctuations of the electronic clouds.
The dispersion interaction is proportional to the ionic polarizability which, in turn, is proportional to the 
ionic volume.  We will, therefore, model the dispersion potential to be proportional to the relative ionic polarizability and 
to the ionic volume exposed to oil~\cite{DoLe12},
\begin{eqnarray}
\label{edis}
\beta U_{d}(z)=\left\{
\begin{array}{l}
 0 \text{ for } z \ge  a  \ , \\
 A_{eff} \alpha [1 - \\
\dfrac{(z/a + 1)^2(2 - z/a)}{4} ] \text{ for } -a<z<a \ ,
\end{array}
\right.
\end{eqnarray}
where $A_{eff}$ is the effective Hamaker constant, which can be adjusted to fit the interfacial tension of one of
the electroltye solutions.  

We are not aware of experimental data for interfacial tensions of sodium salts. The only data available to us is for potassium salts.  Therefore, we must first recalibrate the effective radius of cation.  We find that using the  hydrated radius  of \ce{K+} to be $a=2$~\AA, we obtain a good agreement with the experimental data for \ce{KCl}, see Fig.~\ref{fig_salt_oil}. We will use the same radius of \ce{K+} in  all 
other calculations. We, next, adjusted the value of the effective Hamaker constant.  We find that $A_{eff}=-4$ 
yields a good fit of the excess interfacial tension of \ce{KI} solution, see Fig.~\ref{fig_salt_oil}. 
The same value of $A_{eff}$ is 
then used to calculate the interfacial tensions of other salts containing chaotropic anions, see Fig.~\ref{fig_salt_oil}. The value of $A_{eff}=-4$ is surprisingly close to a theoretical estimate~\cite{DoLe12}, $A_{eff} \approx -4.4$.


\begin{figure}[t]
\vspace{0.2cm}
\includegraphics[width=8.cm]{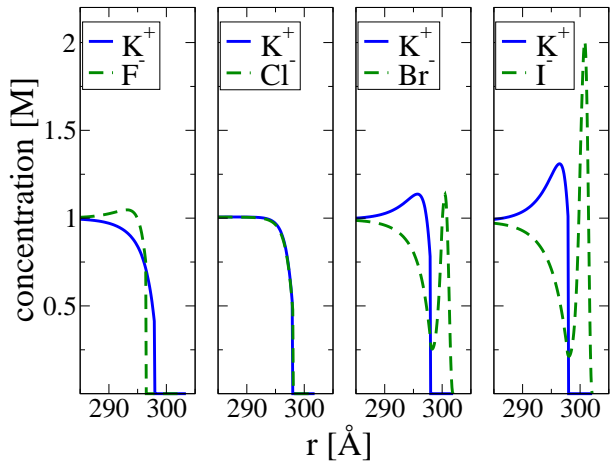}\vspace{0.2cm}
\caption{Ionic density profiles for potassium salts at $1$~M. The GDS is at $r=300$\AA.}
\label{fig_profile_oil}
\end{figure}
In Fig.~\ref{fig_profile_oil}, we show the ionic density profiles.  As expected, 
the dispersion interaction leads to a significant increase in anionic adsorption at a hydrophobic surface, compared to the air-water interface.
\begin{figure}[h]
\vspace{0.2cm}
\includegraphics[width=8.cm]{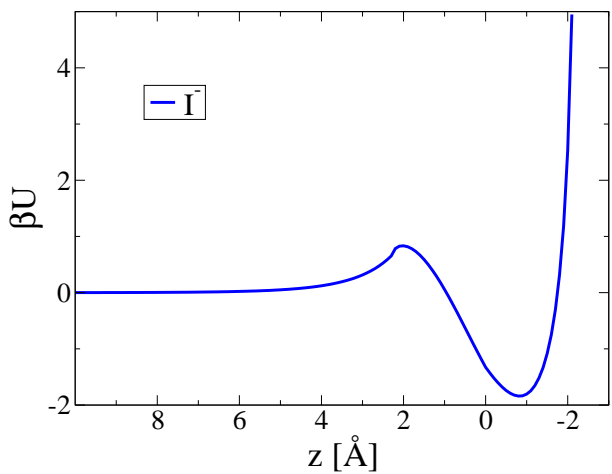}\vspace{0.2cm}
\caption{Oil-electrolyte interface interaction potential of \ce{I-}, at $1$~M.}
\label{fig_potential_oil}
\end{figure}
In Fig.~\ref{fig_potential_oil}, we plot the ion-interface interaction potential of \ce{I-}, which now shows a global minimum
of about $2 k_B T$ at the interface.
Finally, in the  Table~\ref{tab4}, we present the excess electrostatic potential difference across the oil-water interface for various electrolyte solutions at $1$~M concentration.
\begin{table}[h]
\centering
\setlength{\belowcaptionskip}{10pt}
\caption{Surface potential differences for various potassium salts at 1M concentration.}
      \begin{tabular}{|c|c|}
      \hline
      \hline
      Salts         &  calculated [mV]   \\
      \hline
      \hline
      KF           &        $5.85$      \\
      \hline      
      KCl           &        $-0.23$      \\
      \hline
      KNO$_3$       &        $-13.18$     \\
      \hline
      KBr           &        $-17.45$     \\
      \hline
      KI            &        $-30.56$     \\
      \hline
   \end{tabular}
   \label{tab4}
\end{table}

\section{Conclusions}

We have used the PA-DCT to explore the interaction of ions with hydrophobic surfaces.  
The theory shows that ions must
be divided into two classes:  structure making kosmotropes and structure breaking chaotropes.  
In the context of the present theory, structure
making/breaking does not refer to any long-range influence of ions on water, instead
the kosmotropic/chaotropic classification is only used to characterize ionic hydration near a hydrophobic surface.
Ions which have positive JD viscosity $B$ coefficients and have  
historically been called structure makers (kosmotropes) are found to remain strongly hydrated near a hydrophobic surface.  
On the other hand, ions with negative JD viscosity $B$ coefficients, structure breakers (chaotropes),
are found to loose their hydration sheath and as the result of their large polarizability can become
adsorbed to the interface.  The theory shows that ionic polarizability is an essential ingredient for the adsorption of
chaotropic anions. The huge cost in electrostatic solvation free energy prevents adsorption of ions of low polarizability  at hydrophobic interfaces.  A small adsorption~\cite{HoHe09,NeHo12} of  non-polarizable ions observed in the recent classical explicit water simulations has been attributed to the artificial electrostatic surface potential of the neat air-water interface which exists in point charge water  models~\cite{BaSt12}.  The same artificial surface potential of classical water models 
was shown in the present  Review to lead to an 
excessive adsorption of the polarizable ions in the  
PFFMD simulations.  The objective of 
future work should be development of classical water
models with very low electrostatic surface potential.  Such water models should then
be compatible with the polarizable force fields and could then be used to study the interaction
of ions with more complicated hydrophobic surfaces and proteins.

\section{Acknowledgments}
This work was partially supported by the CNPq, FAPERGS, FAPESC, INCT-FCx, and by the US-AFOSR under the grant 
FA9550-12-1-0438.
\bibliography{ref.bib}

\end{document}